\documentclass[aps,prl,twocolumn,groupedaddress,showpacs,floatfix]{revtex4}
\usepackage{amssymb}
\usepackage{epsf,citesort,rotate,setspace,afterpage,graphicx,subfigure}

\begin{document}

\title{On the Origin of Power-Law Fluctuations in Stock Prices}
\author{Vasiliki Plerou$^1$, Parameswaran Gopikrishnan$^{1,2}$, Xavier Gabaix%
$^3$, and H. Eugene Stanley$^1$}

\affiliation{$^1$ Center for Polymer Studies and Department of Physics,
Boston University, Boston Massachusetts 02215.\\ $^2$ Goldman Sachs
\& Co., New York, New York 10004.\\ $^3$ Economics Department, MIT, and NBER,
Cambridge Massachusetts 02142 \\ }

\date{Published in Quantitative Finance {\bf 4} (February 2004) C11-C15}

\begin{abstract}
We respond to the issues discussed by Farmer and Lillo (FL) related to our
proposed approach to understanding the origin of power-law distributions in
stock price fluctuations. First, we extend our previous analysis to 1000 US
stocks and perform a new estimation of market impact that accounts for
splitting of large orders and potential autocorrelations in the trade flow.
Our new analysis shows clearly that price impact and volume are related by a
square-root functional form of market impact for large volumes, in contrast
to the claim of FL that this relationship increases as a power law with a
smaller exponent. Since large orders are usually executed by splitting into
smaller size trades, procedures used by FL give a downward bias for this
power law exponent. Second, FL analyze 3 stocks traded on the London Stock
Exchange, and solely on this basis they claim that the distribution of
transaction volumes do not have a power-law tail for the London Stock
Exchange. We perform new empirical analysis on transaction data for the 262
largest stocks listed in the London Stock Exchange, and find that the
distribution of volume decays as a power-law with an exponent $\approx 3/2$
--- in sharp contrast to FL's claim that the distribution of transaction
volume does not have a power-law tail. Our exponent estimate of $\approx 3/2$
is consistent with our previous results from the New York and Paris Stock
Exchanges. We conclude that the available empirical evidence is consistent
with our hypothesis on the origin of power-law fluctuations in stock prices.
\end{abstract}

\maketitle

\draft


We recently proposed a testable theory for the origin of the
empirically-observed power-law distributions of financial market variables
such as stock returns, volumes, and frequency of trades~\cite{Gabaix03}. Our
theory explains the power-law exponent of the distribution of returns by
deriving a square-root functional form for market impact (``square-root
law'') that relates price impact and order size. Our previous empirical
analysis gave results that support the square-root law of market impact.

Farmer and Lillo (FL)~\cite{Farmer03} raise some issues related to the
empirical validity of the theory proposed in Ref.~\cite{Gabaix03}. Their
discussion is based on the following arguments:

\begin{enumerate}
\item FL claim that the price impact function grows slower than a
square-root law. Interestingly, FL's empirical analysis does find a
power-law relationship for market impact with exponent $\beta \approx 1/2$
for volumes smaller than a threshold, consistent with the square-root form
of market impact in our theory. However, for large volumes FL claims $\beta
<0.2$ [for the New York Stock Exchange (NYSE)] and $\beta \approx 0.26$
[based on analyzing 3 stocks in the London Stock Exchange (LSE)]. FL do not
compute error-bars for $\beta $ for large volumes, but claim that, from a
visual comparison, $\beta =1/2$ (square-root law) is inconsistent with the
data.

\item FL argue that the empirical analysis that we presented in support of
the square-root functional form of market impact \cite{Farmer03} is
``invalidated'' by the ``long-memory nature of order flow.''

\item FL analyze the volume distribution of 3 stocks from the London Stock
Exchange and claim that the volume distribution does not follow a power-law.
Consequently, FL conclude that volume fluctuations do not determine the
power-law tail of returns.
\end{enumerate}

We first outline our responses to these criticisms and then present our
detailed response with results of our new analysis.

\begin{enumerate}
\item FL find $\beta < 1/2$ for large volumes from analyzing the average
value of return for a trade for a given trade size. FL's procedure for
estimating price impact is flawed since large orders are usually executed by
splitting into smaller size trades, so the procedure used by FL gives a
downward bias for the power law exponent $\beta$ defined in our theory~\cite{Gabaix03,Gabaix04}, giving rise to an apparent exponent value $%
\beta^{\prime}$ smaller than the correct value $\beta$. In fact FL's
procedure gives $\beta < 1/2$ for large volumes---precisely the domain in
which we expect the order-splitting effect to be dominant---and therefore a
downward bias for $\beta$.

\item Although we present new estimators to address this point, we
believe FL's argument to be incorrect since long-memory in order flow
clearly does not imply the same for returns, so FL's criticisms about
our estimation procedure do not seem relevent. To address a potential
problem of long memory in order flow, we draw from a new estimator for
measuring market impact \cite{Plerou04} and extend our previous
analysis to the 1000 largest NYSE stocks. Our new estimation confirms
that the market impact function does behave as a square-root function
of the volume.

\item We analyze 262 largest stocks listed in the London Stock Exchange. Our
analysis of volume distribution for these 262 stocks shows that the
distribution of volume decays as a power-law with an exponent $\approx 3/2$
in agreement with our previous results for the NYSE and the Paris Bourse. In
fact our analysis of the volume distribution for the same stocks analyzed by
FL shows a clear power-law behavior with exponent $\approx 3/2$ --- in
contrast to FL's claim.
\end{enumerate}

Define\ $S_{i}$ as the price of the stock after trade $i$, 
\begin{equation}
\delta p_{i}\equiv \log S_{i}-\log S_{i-1}  \label{e1x}
\end{equation}%
as the return concomitant to trade $i$, so the return over a fixed time
interval $\Delta t$ is 
\begin{equation}
r\equiv \sum_{i=1}^{N}\delta p_{i},  \label{e2x}
\end{equation}%
where $N$ is the number of trades in $\Delta t$. Let $q_{i}$ be the number
of shares traded in trade $i$ so 
\begin{equation}
Q\equiv \sum_{i=1}^{N}q_{i}  \label{e3x}
\end{equation}%
is the total volume in interval $\Delta t$. We define the trade imbalance 
\begin{equation}
\Omega \equiv \sum_{i=1}^{N}\epsilon _{i}q_{i}  \label{e4x}
\end{equation}%
where $\epsilon _{i}=1$ indicates a buyer-initiated trade and $\epsilon =-1$
denotes a seller initiated trade. We denote by $V$ the size of a large
order, which can be executed in several trades.

\textbf{1. Measuring market impact}

Let $\Delta p$ be the change in price caused by a large order of size $V$,
all else remaining the same. Our theoretical approach\cite{Gabaix03} derives
a power-law functional form for the market impact function~\cite{noteimpactfn}, 
\begin{equation}
\Delta p\sim V^{\beta } \,.  \label{DpDef1}
\end{equation}
We hypothesized $\beta =0.5$ and supported the hypothesis by empirical
analysis.

Our hypothesis Eq.~(\ref{DpDef1}) pertains to $\Delta p$, the total
impact in price of a large order of size $V$. In practice, as outlined
in Ref.~\cite{Gabaix03}, large orders are executed by splitting into
orders of smaller size which are observed in the trade time series as
the trade size $q_{i}$.  The empirical analysis of
Refs.~\cite{Farmer03} and \cite{Lillo03} refer to the relationship
local price change $\mathbf{E}(\delta p|q)$ and not the price impact
$\mathbf{E}(\Delta p|V)$ that we are interested in. The true market
impact function $\mathbf{E}(\Delta p|V)$ is indeed notoriously
difficult to measure since the information about the unsplit order
size is usually proprietary and not available, neither in our data nor
in the data analyzed by FL.

FL claim that the price impact function grows more slowly than a square-root
function for large volumes. The basis for their claim is the analysis
presented in Ref.~\cite{Lillo03} that $\mathbf{E}(\delta p|q)\sim q^{\beta }$
with $\beta =0.5$ for small $q$ and $\beta =0.2$ for larger $q$. While $%
\mathbf{E}(\delta p|q)$ indeed grows less rapidly than a square-root, as
reported in Ref.~\cite{Plerou02}, $\mathbf{E}(\delta p|q)$ neither
quantifies price impact of large trades, nor does it contradict our theory
and empirical results~\cite{Gabaix03}. This is because a trade by trade
analysis of $\mathbf{E}(\delta p|q)$ leads to a \textit{biased\/}
measurement of full price impact and the exponent $\beta $, since it does
not take into account the splitting of trades~\cite{Gabaix03,Gabaix04}.

Consider an example. Suppose that a large fund wants to buy a large number $V
$ of shares of a stock whose price is \$100. The fund's dealer may offer
this large volume for a price of \$101. Before this transaction, however,
the dealer must buy the shares. The dealer will often do that progressively
in many steps, say 10 in this example. In the first step, the dealer will
buy $V/10$ shares, and the price will go say, from \$100 to \$100.1, and in
the second the price will go from \$100.1 to \$100.2. After some time
elapses, the price will have gone to \$101 in increments of \$0.1. At this
stage, the dealer has his required number of shares, and hands them over to
the fund manager at a price of \$101. The true price impact here is 1\%,
since the price has gone from \$100 to \$101. But in any given transaction,
the price has moved by no more than \$0.1. So Ref.~\cite{Farmer03} would
find an ``apparent'' price impact of no more than \$0.1, i.e. 0.1\% of the
price. Since as the transaction is executed the price of the stocks goes
from \$100 to \$101, the true price impact is 1\%. As a result the procedure
of FL will measure a value 10 times smaller than the true value. This
downward bias explains why FL find in Fig.~2 a maximum impact of 0.1\%---a
very small price impact. Other evidence in economics \cite{Chan95,Keim97}
finds impacts that are up to 40 times larger than that of FL's analysis.
Likewise our evidence pertains to large impacts, captured by Fig.~2 of our
paper which shows on the vertical axis values of $r^{2}$ equal up to $200$
times the variance $\sigma ^{2}$, i.e., values of return $r$ up to $\sqrt{200%
}\approx 14$ standard deviations.

We can quantify the bias in the above example. Suppose that a trade of size $%
V$ is split into $K=V^{\alpha }$ (10 in our example) trades of equal size $%
q=V/K=V^{1-\alpha }$, with $0<\alpha <1$. Then the apparent impact $\delta p$
incurred by each trade (0.1 $\%$ in our example) will be $1/K$ (1/10 in our
example)\ of the total price impact $V^{\beta }$ (1\% in our example), i.e. $%
\delta p=V^{\beta }/K$ $=$ $V^{\beta -\alpha }$. So a power law fit of $%
\delta p$ vs $q$, such as the one presented in Fig. 2 of FL, will give $%
\delta p\sim q^{\beta ^{\prime }}$ with~\cite{footnoteBetaLessThan1} 
\[
\beta ^{\prime }=\left( \beta -\alpha \right) /\left( 1-\alpha \right)
<\beta .
\]%
The \textquotedblleft trade by trade\textquotedblright\ measurement of the
price impact, as performed by Ref.~\cite{Farmer03,Lillo03}, leads to a
biased measurement $\beta ^{\prime }$ of the exponent $\beta $ of the true
price impact.

It is to address this bias that we examine $\mathbf{E}(r^{2}|Q)$ in Ref.~%
\cite{Gabaix03}. As is well established empirically, the sign of returns is
unpredictable in the short term, so the reasoning in Ref.~\cite{Gabaix04}
shows that $\mathbf{E}(r^{2}|Q)$ will not be biased~ \cite{notetmp}.

Our analysis \cite{Gabaix03} was presented with data for the 116 most
actively traded stocks. To check if the result of $\beta =0.2$ for large
volumes presented in Refs.~\cite{Farmer03} and \cite{Lillo03} could arise
from increasing the size of the database, we now extend our analysis to the
1000 largest stocks in our database for the 2-yr period 1994-95. Figure~\ref%
{qdens}(a) confirms that $\mathbf{E}(r^{2}|Q)\sim Q$ as predicted by our
theory.

\begin{figure}[hbt]
\centering      
 \includegraphics[width=0.4\textwidth,height=0.4%
\textwidth,angle=-90]{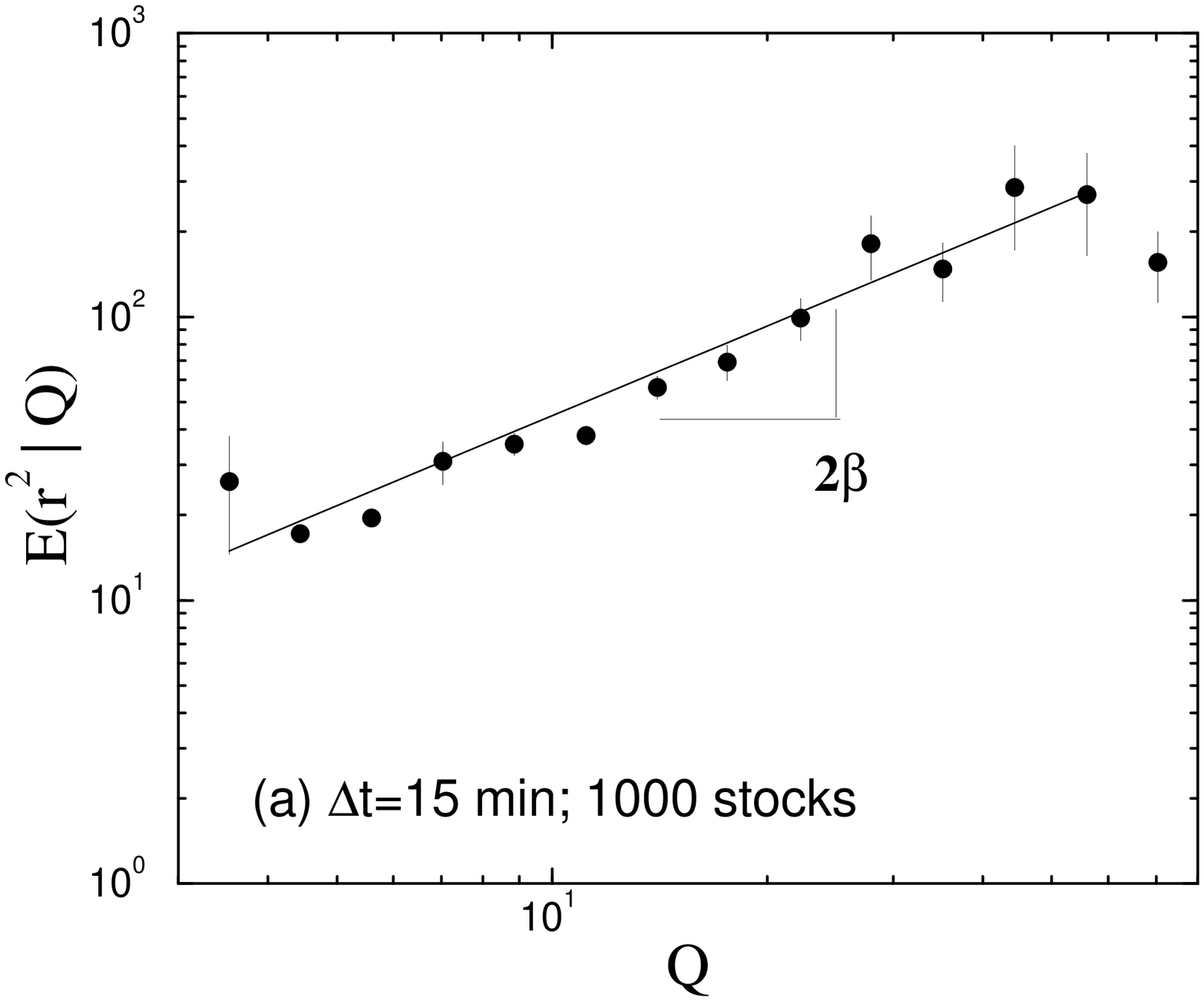}
\par
\includegraphics[width=0.4\textwidth,height=0.4%
\textwidth,angle=-90]{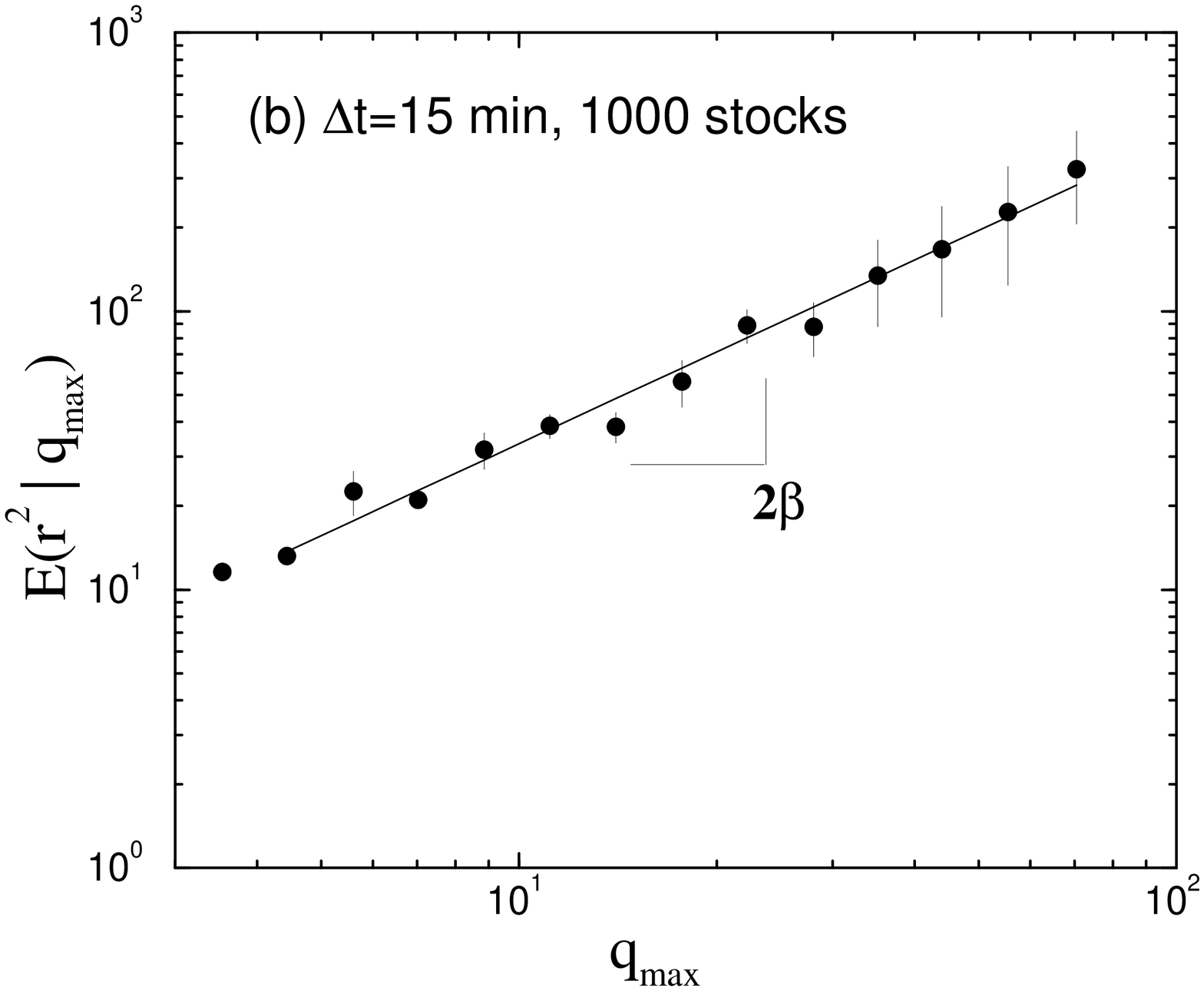}
\end{figure}
\begin{figure}[hbt]
\centering      \includegraphics[width=0.4\textwidth,height=0.4%
\textwidth,angle=-90]{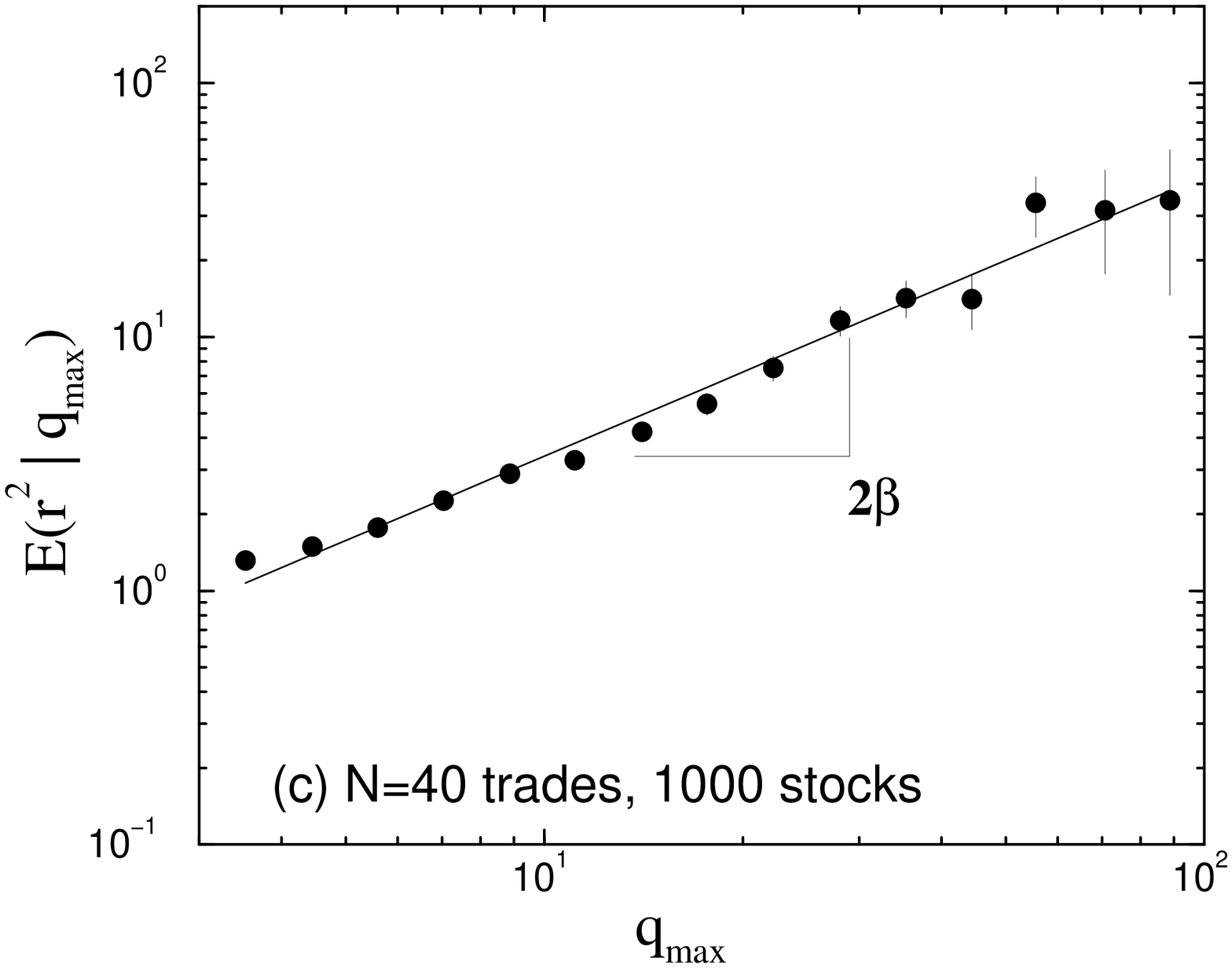} 
\label{qdens}
\caption{(a) Conditional expectation function $\mathbf{E}(r^{2}|Q)$ of the
squared return for a given volume for $\Delta t=$~15 min for 1000
largest stocks in the NYSE, Amex and NASDAQ for the period 1994-95. We
have normalized $r$ for each stock to zero mean and unit variance, and
$Q$ is normalized by its first centered moment. This normalization
procedure allows for a data collapse for different stocks and the plot
represents an average for 1000 stocks. Regressions in the range
$3<Q<70$ give values of the exponent $\protect\beta =1.05\pm
0.08$. (b) Conditional expectation function $\mathbf{E}\left(
r^{2}|q_{\mathrm{max}}\right) $ of the squared return for a given
$q_{\mathrm{max}}$ for $\Delta t=15$~min. Here $q_{\mathrm{max}}$ is
the largest trade size in the$15$~min interval. Power-law regression
gives the value of the exponent $2\protect\beta =1.09\pm 0.06$
consistent with $%
\protect\beta =0.5$. (c) Conditional expectation function $\mathbf{E}\left(
r^{2}|q_{\mathrm{max}}\right) $ of the squared return for a given $q_{%
\mathrm{max}}$ for fixed number of trades $N=40$. Power-law regression gives
the value of the exponent $2\protect\beta =1.10\pm 0.06$. Source: Ref.  
\protect\cite{Plerou04}.}
\end{figure}

\textbf{2. Robustness of estimation against long-memory of order flow}

FL argue that the empirical analysis that we presented in support of
the square-root functional form of market impact is \textquotedblleft
invalidated\textquotedblright\ by the \textquotedblleft long-memory
nature of order flow\textquotedblright . FL's argument is based
entirely on the \textit{assumption\/} that returns due to each
transaction $i$ can be written as $r_{i}=\epsilon _{i}q_{i}^{\beta }$
where $\epsilon _{i}=1$ for a buy trade and $\epsilon _{i}=-1$ for a
sell trade. Under this assumption, FL then argue that our estimator
$\mathbf{E}(r^{2}|Q)$ is affected by the long-range correlations in
the trade signs $\epsilon _{i}$~\cite{Farmer03,Bouchaud04}. FL give
some numerical evidence for this potential effect for small to
moderate volumes.

\begin{figure}[hbt]
\centering   \includegraphics[width=0.4\textwidth,height=0.4%
\textwidth,angle=-90]{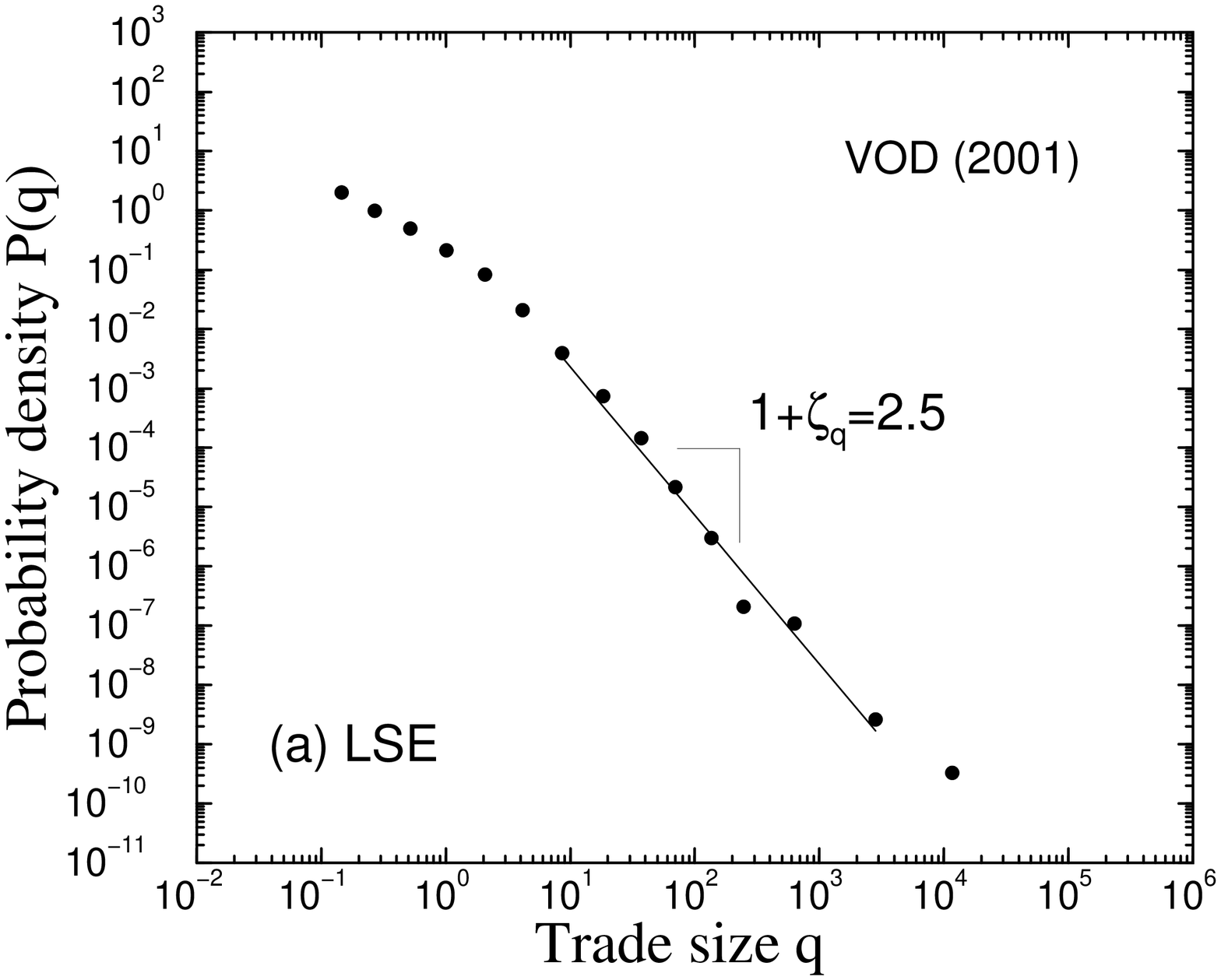}
\par
\includegraphics[width=0.4\textwidth,height=0.4%
\textwidth,angle=-90]{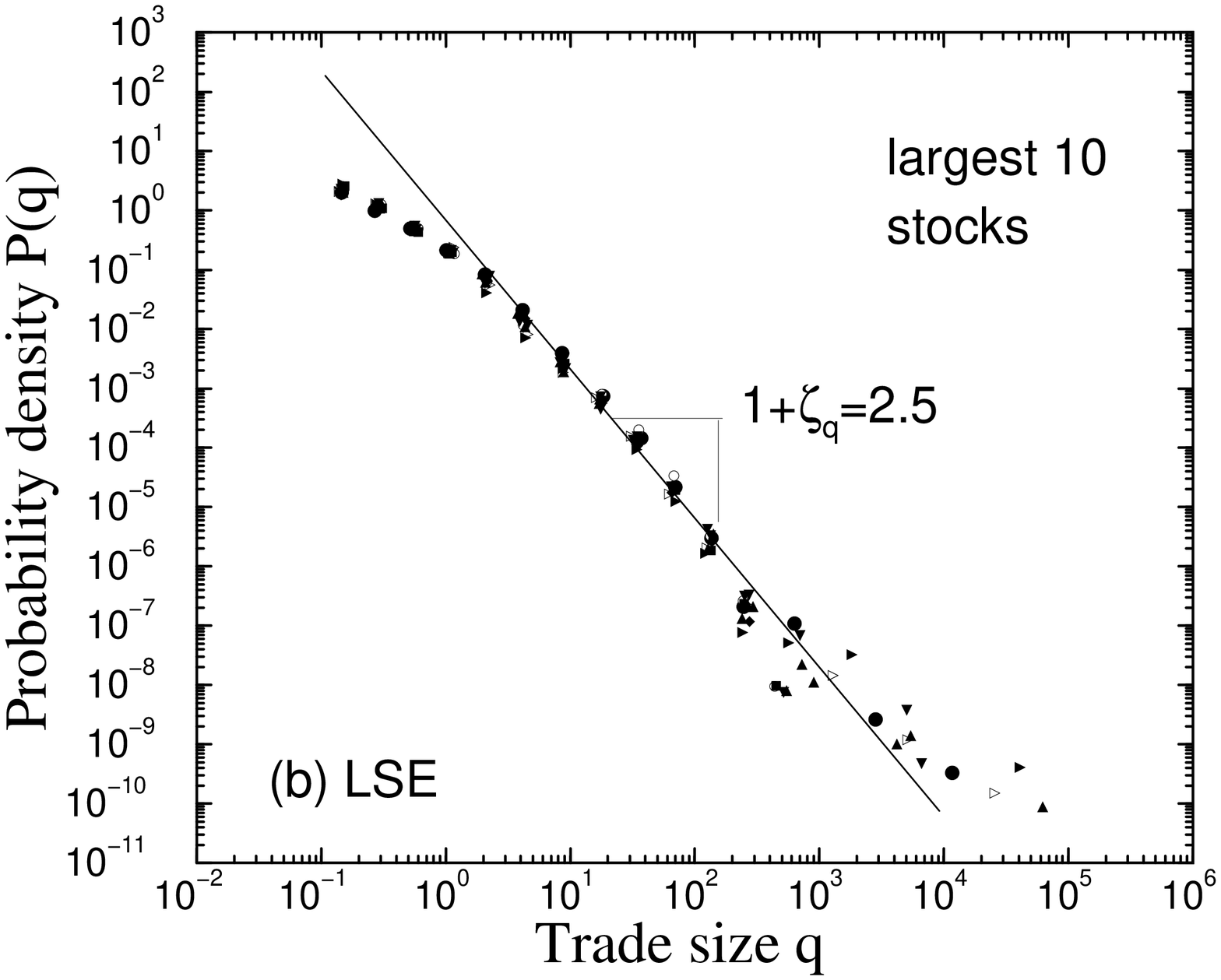}
\end{figure}

All of the tests shown by FL are for a \textit{fictitious} return
$f_{i}$ constructed on a trade-by-trade basis as $f_{i}=\epsilon
_{i}q_{i}^{\beta }$. FL's argument and the tests shown in Fig.~1 of FL
are for the fictitious return. In reality, return can certainly not be
expressed as $r_{i}=\epsilon _{i}q_{i}^{\beta }$, with $\epsilon _{i}$
being the trade indicator. Indeed, if this were true, returns
themselves would be long-range correlated---a possibility long known
to be at odds with empirical data. Since the sign of the return
$r_{i}$ and that of the trade sign $\epsilon _{i}$ are clearly not
equal, FL's argument about our estimation procedure being affected by
the long-memory nature of the trade sign ($\epsilon _{i}$) is
incorrect (See Ref.~\cite{Bouchaud04} on a related point).

\begin{figure}[hbt]
\centering   \includegraphics[width=0.4\textwidth,height=0.4%
\textwidth,angle=-90]{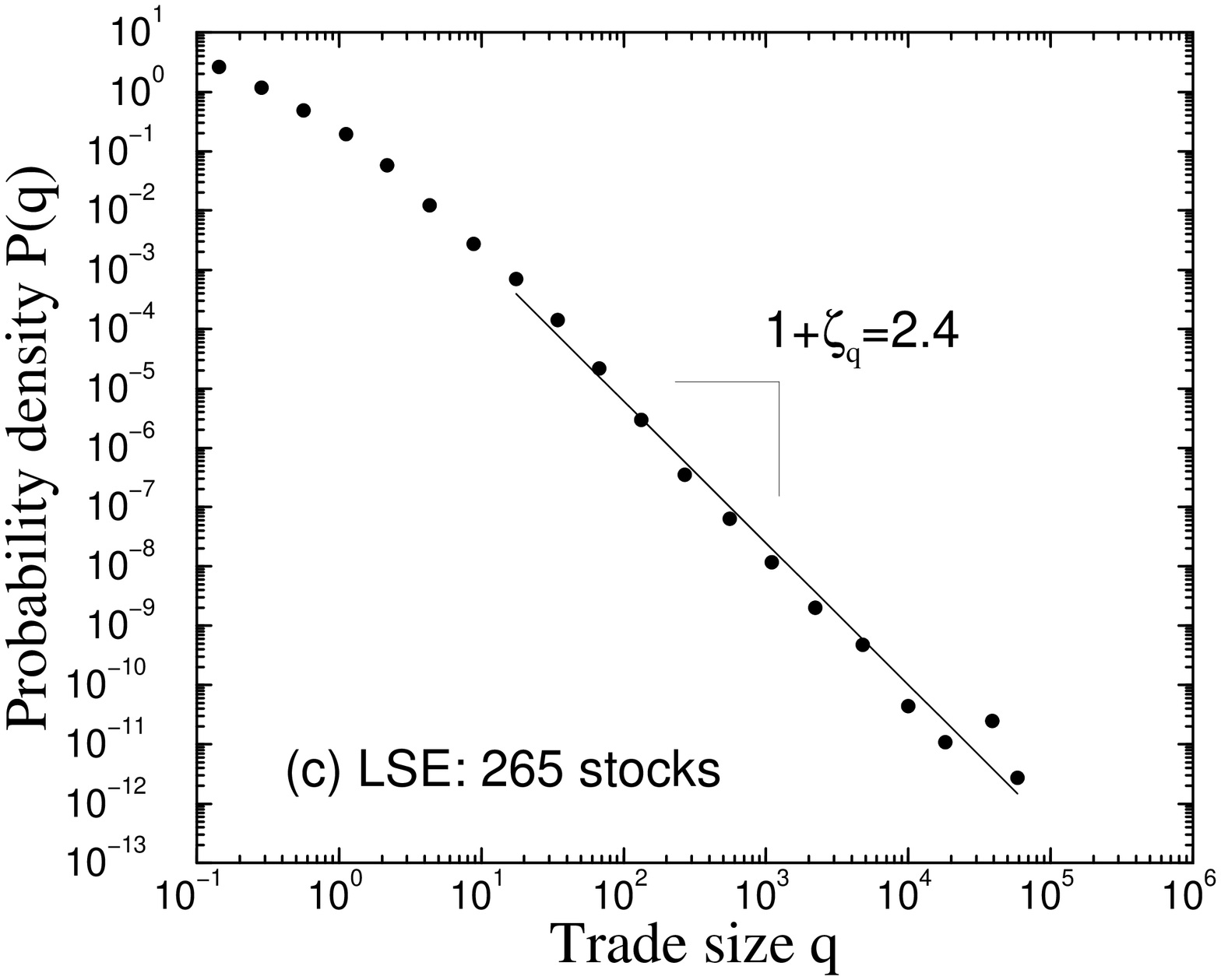}
\par
\end{figure}

\begin{figure}[hbt]
\includegraphics[width=0.4\textwidth,height=0.4%
\textwidth,angle=-90]{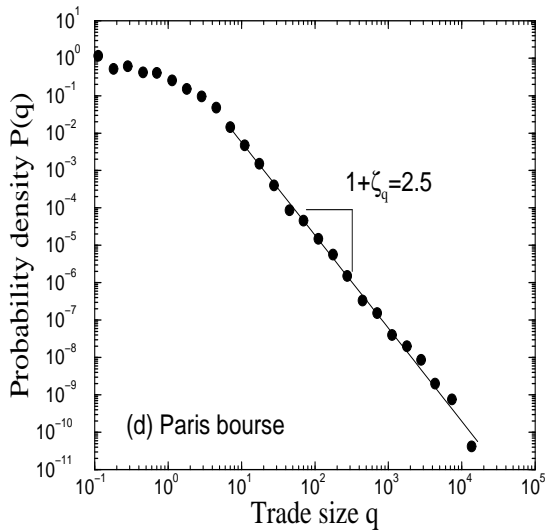}
\par
\caption{ (a) Probability density function of trade volumes for
Vodafone Inc. (VOD) for 2001. A power-law fit in the region
$10<q<1000$ gives value of the exponent $\protect\zeta _{q}=1.5\pm
0.1$. In contrast FL finds much more rapid decay. (b) Probability
density function of trade volumes for 10 largest stocks listed in the
London Stock Exchange, showing clear evidence of power-law decay with
exponent $\protect\zeta _{q}=$ $1.5\pm 0.1$, consistent with our
previous results \protect\cite{Gopi00} for the New York Stock Exchange
and for the Paris Bourse. Here $q$ are normalized by its first
centered moment, so all 10 distributions collapse on one curve. We
find an average exponent $\protect\zeta _{q}=$1.59 $\pm 0.09$. (c)
Same as (b) for all 265 stocks in our sample where $q$ is normalized
for each stock by its first centered moment. We find $\protect\zeta
_{q}=$ $1.4\pm 0.09$. (d) Probability density function of trade
volumes for 30 largest stocks listed in the Paris Bourse obtained by
the same procedure. We find $\protect\zeta_{q}=$ $1.49\pm
0.03$~\protect\cite{Gabaix04}.}
\label{qlse}
\end{figure}

Although FL's argument is incorrect, to address the more general
concern that the autocorrelations of the trade signs $\epsilon _{i}$
might bias our analysis, we draw from a forthcoming paper
\cite{Plerou04}, which performs the following
analysis~\cite{footnoteSplitMatters}. For each intervl $
\Delta t$ define $q_{\mathrm{max}}$ as the size of the  largest trade. In
our theory, if the largest trade $V_{\mathrm{max}}$ is large, it will
have a the major influence on the value of return, so that one will
have $ r^{2}\sim V_{\mathrm{max}}^{2\beta }\sim
q_{\mathrm{max}}^{2\beta }$. Hence $ q_{\mathrm{max}}$ gives us a
diagnostic value of the behavior of the largest trade, independently
of a potential collective behavior. We detail this in
Ref. \cite{Plerou04}. We compute $\mathbf{E}(r^{2}|q_{\mathrm{max}})$
and find [Fig.~\ref{qdens}(b)].

\begin{equation}
\mathbf{E}(r^{2}|q_{\mathrm{max}})\sim q_{\mathrm{max}}.  \label{r2cqmax}
\end{equation}

In addition to the above, to ensure that our estimation is robust to varying
number of trades in a fixed $\Delta t$, we have computed $\mathbf{E}%
(r^{2}|q_{\mathrm{max}})$ for fixed number of trades instead. Figure~\ref%
{qdens}(c) shows that $\mathbf{E}(r^{2}|q_{\mathrm{max}}) \sim q_{\mathrm{max%
}}$ for $r$ over $N=40$ trades.

We would like to emphasize that in Fig.~\ref{qdens} we consider very large
trades that are up to 70 times the first moment of volume. They correspond
to returns of up to 14 standard deviations of returns. This confirms that we
study very large trades and returns---the ones that are relevant for the
study of power law fluctuations, while in contrast FL's analysis does not
systematically treat large trades. 

We conclude that the procedure used in Ref.~\cite{Farmer03} has a downward
bias of the price impact $\beta $ of large trades. When we perform more
appropriate analysis, we confirm that the $\beta \approx 1/2$. This
corroborates our hypothesis~\cite{Gabaix03,Gabaix04} that large fluctuations
in volume cause large fluctuations of prices.

\textbf{3. Half-cubic power-law distribution of volumes}

The last claim of FL pertains to the very nature of the volume distribution.
They present the results of their analysis of three stocks in the London
Stock Exchange and claim their analysis shows no evidence for a power-law
distribution.

We analyze the same database which records \textit{all\/} trades for \textit{%
all\/} stocks listed in the London Stock Exchange. From this database, we
first examine one stock -- Vodafone, VOD --- which is analyzed by FL. For
this stock, we compute the volume distribution and find clear evidence for a
power-law decay [Fig.~\ref{qlse}(a)] 
\begin{equation}
P(q)\sim q^{-\zeta _{q}-1}  \label{densq}
\end{equation}%
with exponent $\zeta _{q}=1.5\pm 0.1$, in agreement with our previous
results for the NYSE and the Paris Bourse~\cite{Gopi00,Gabaix03}, but in
sharp contrast to the FL results who claim a thin-tailed distribution for
the same data.

For the 10 largest stocks in our sample, Fig.~\ref{qlse}(b) shows that
$P(q)$ is consistent with the same power-law of $\zeta_q = 3/2$
---consistent with our earlier finding for the NYSE~\cite{Gopi00}. We
extend our analysis to the 265 largest LSE stocks and find similar
results [Fig.~\ref{qlse}(c)].

To test the universal nature of this distribution, we analyze data for 30
largest stocks listed in the Paris Bourse and find that $P(q)$ is consistent
with the a power-law with almost identical exponents $\zeta _{q}=3/2$ [Fig.~%
\ref{qlse}(d)].

In summary, the analysis of Ref.~\cite{Farmer03} pertains to small to
moderate trades. FL's estimation is biased for large trades, so FL can
detect only very small price impacts, less than 0.1\%. When we use our
more general procedure and study significantly larger data, we confirm
our initial finding of a square root price impact function.  We
conclude that the available evidence is consistent with our hypothesis
\cite{Gabaix03,Gabaix04} that large fluctuations the volume traded by
large market participants may contribute significantly to the large
fluctuations in stock prices~\cite{Plerou04}.

\medskip Acknowledgments. We thank the NSF's economics program for financial
support, and Joel Hasbrouck, Carl Hopman, Gideon Saar, Jeff Wurgler and
especially Gilles Zumbach for helpful conversations and help with obtaining
the data analyzed.

{\it \bf Note added in press:} After our initial submission, it has
become clear that FL's claim of a non-power-law distribution of trade
sizes [2] for the LSE stocks is based on incomplete data. FL's
analysis excludes the upstairs market~\cite{blockTrades} trades which
contain the largest trades in the LSE. In contrast, our result of a
3/2 power-law exponent for the volume distribution is based on data
containing all trades (both the upstairs and the downstairs market
trades) in the LSE. By excluding the large trades in the upstairs
market, FL set an artificial truncation at large volume, so FL's
finding of a non-power-law distribution of volume is merely a trivial
artifact of incomplete data. Although FL claim in their note added
that ``it has been shown that large price fluctuations in the NYSE
(including the upstairs market) and the electronic portion of the LSE
are driven by fluctuations in liquidity'' their new analysis and
findings are tainted by the same problems as in their present comment:
(i) incompleteness (absence of the upstairs market trades) of the data
analyzed and (ii) they do not take into account the splitting of large
orders.

Gabaix et al.~\cite{Gabaix03,Gabaix04} and FL~\cite{Farmer03} discuss
two distinct possibilities respectively: (i) large price changes arise
from large trades and (ii) large price changes arise from fluctuations
in liquidity~\cite{Plerou02,Plerou00}.  While we believe that both
mechanisms play a role in determining the statistics of price changes,
our empirical findings support the possibility that the specific
power-law form of the return distribution arises from large trades.



\begin{references}
\bibitem{footnoteBetaLessThan1}
The inequality below holds if $\beta <1$. There is wide agreement that $\beta $
should be no greater than $1/2$. This is because the power law exponent of
returns $\zeta _{r}$
and the power law exponent of volumes satisfy $\zeta _{r}\leq \zeta
_{V}/\beta $ \cite{Gabaix03}, so that the empirical values $\zeta
_{V}\approx \zeta _{r}/2\approx 3/2$ imply $\beta $ $\precsim 1/2$.
\bibitem{footnoteSplitMatters}
We are detailing the foundations of this analysis in Ref.~\cite{Plerou04}.
The specifics of the split matter in principle. In Ref. \cite{Gabaix03} we
present a theory of power law splitting in which the size of the largest
chunk, $q_{\max }$, is proportinal to the size of the entire order, $V$.
\bibitem{noteimpactfn}
As in Ref.~\cite{Gabaix03}, we interpret $\Delta p\sim V^{\beta }$ to
mean that there exist a slowly varying function $L\left( V\right) $
such that for large $V$, $\Delta p/V^{\beta }L\left( V\right)
\rightarrow 1$. A function is called slowly varying when for all
$t>0$, $\lim_{V\rightarrow \infty }L\left( tV\right) /L\left( V\right)
=1$. Typical slowly varying functions are $L\left( V\right) =a$, or
logarithmic corrections: $L\left( V\right) =a\ln \left( V\right)
^{\alpha }$, where $a$ and $\alpha $ are constants.
\bibitem{blockTrades} Large trades tend to be executed in the ``upstairs'' market by
bilateral arrangements than through the order book.
\end{references}
\end{document}